\documentclass{aa}
\usepackage{graphicx,subfigure,rotating}

\usepackage{natbib,references}
\bibpunct{(}{)}{;}{a}{}{,} 

\begin{document}
   \title{Doppler Images of the RS CVn Binary HR 1099 (V711 Tau) from the MUSICOS 1998 Campaign\thanks{Based on observations obtained during the MUSICOS 98 
   MUlti-SIte COntinuous Spectroscopic campaign from Kitt Peak National Observatory, USA, and  Mt. Stromlo Observatory, Australia.}}


\author{D. Garc\'{\i}a-Alvarez\inst{1}
          \and J.R. Barnes\inst{2}
	  \and A. Collier Cameron\inst{2}
	  \and J.G. Doyle\inst{1}\\
	  \and S. Messina\inst{3}
	  \and A. F. Lanza\inst{3}	  
	  \and M. Rodon\`o\inst{4}
}
          
\offprints{D. Garc\'{\i}a-Alvarez\\
\email{dga@star.arm.ac.uk}}

\institute{Armagh Observatory, College Hill, Armagh BT61 9DG N.Ireland
	\and School of Physics and Astronomy, University of St Andrews, St Andrews, Fife KY16 9SS, Scotland
	\and Catania Astrophysical Observatory of the National Institute for Astrophysics, Via S. Sofia, 78 I 95123 Catania, Italy
	\and Department of Physics and Astronomy, Catania University,  Via S. Sofia, 78  I-95123 Catania, Italy
}
   \date{Received, 2002 / accepted, 2003}

   \abstract{We present Doppler Images of the RS CVn binary system HR 1099 (V711 Tau) from spectra taken in two different sites, KPNO and MSO, during the MUSICOS 1998 campaign. Contemporaneous APT photometry is used to constrain the Doppler Images. The resulting maximum entropy reconstructions based on the least-squares deconvolved profiles, derived from $\sim$2000 photospheric absorption lines, reveal the presence of starspots at medium-high latitudes. We have obtained maps for both components of the binary system for the first time. The predominant structure in the primary component is an off-centered polar spot, confirming previous works on the same target by using independent codes. The result is verified by using both data sets independently. The lower spectral resolution data set gives a less detailed map for the MSO data set. The images obtained for the secondary component show a low latitude spot around orbital phase 0.7. This spot seems to mirror the structure seen on the primary. It might suggest that tidal forces may influence the spot distribution on this binary system. 
\keywords{Stars: binaries: spectroscopic -- Stars: late--type  -- Stars: individual: HR 1099 -- 
Stars: activity -- Stars: imaging -- Stars: starspots}
}

\maketitle

\markboth{D. Garc\'{\i}a-Alvarez et al.: Doppler Images of HR 1099 (V711 Tau)}
{D. Garc\'{\i}a-Alvarez et al.: Doppler Images of HR 1099 (V711 Tau)}

\section{Introduction}
RS CVn binary systems consist of a chromospherically active evolved 
star tidally locked to a main-sequence or sub-giant companion. Short orbital periods ranging from a few days to 20 days are typically observed. The RS CVn high level of activity has been 
measured from radio to hard X-rays and, for fast rotators, it 
approaches the saturation limits for chromospheric, transition region and 
coronal emission. One of the striking aspects of these systems is their propensity to
flare \citep{Doyle92,Garcia-Alvarez02b}. Moreover,  RS CVns show optical photometric variations which are 
believed to arise from the rotational modulation of photospheric spots 
\citep{Rodono86,Byrne95}, large scale versions of dark solar spots, that provide 
evidence for large-scale magnetic fields. \citet{Donati99} and \citet{Donati03} showed that Zeeman Doppler Imaging can also provide evidence of such magnetic structures. Short-period RS CVn-like systems, through their 
rotational modulation, can then provide information on the morphology and 
two-dimensional structure of active regions in stellar atmospheres.

\citet{Deutsch58,Deutsch70} first proposed the use of line profiles as a tool to map stellar surfaces. \citet{Goncharsky77} developed the first inversion technique with regularization, but it was \citet{Vogt83} who first called it Doppler Imaging. The technique exploits the correspondence between the wavelength position across a rotationally
broadened spectral line and the spatial position across the stellar disk at a given rotation phase. \citet{Vogt87} addressed the difficulties in inverting the spectral line profiles into an
image of the stellar disk. Doppler imaging has provided one of the few direct pieces of evidence for the existence of starspots, or localised plages, but the
technique is severely restricted. The data must be of very high quality (S/N $>$ 200) and the method can only be applied to rapidly
rotating stars ($v$\,sin\,$i$ $>$ 20 km\,s$^{-1}$) up to a limit where the signal is swamped by terrestrial and stellar line blends ($v$\,sin\,$i$ $>$ 160 km\,s$^{-1}$).
The stellar inclination is limited to a range of 20$^{\circ}$$<$i$<$90$^{\circ}$, and the physical properties of the atmosphere must also be known,
although the process is fairly insensitive to the detailed physics of spectral line formation. Several reviews, explaining the principles of this technique, can be found in the literature \citep[e.g.,][]{Rice89,Collier-Cameron01,Rice02}.

The target under consideration, HR 1099 (V711 Tau, $03^h36^m47^s$ $+00^o35'16''$  (J2000), { V=5.64, B$-$V=0.92}), is a close 
double-lined spectroscopic binary. At a distance of 29 pc (The {\it{Hipparcos}} and Tycho 
Catalogues \citep{ESA97}), 
HR 1099 is the nearest 
and brightest of the classical RS CVns with a  
K1\,IV primary and a G5 V secondary tidally locked in a 
$2\fd8$ orbit.  \citet{Fekel83} provided the orbital 
parameters of the binary and obtained  masses, radii, 
and spectral types for both components. \citet{Garcia-Alvarez02b} obtained the most recent orbital parameters of the system. The K sub-giant nearly fills its Roche lobe 
and is by far the more active of the two {components} \citep{AyresLinsky82,Robinson96}.
{Evolutionary models 
suggest that mass transfer from the K primary onto the G secondary may begin 
within $10^{7}$ years \citep{Fekel83}.} 
The primary exhibits conspicuous signatures of chromospheric activity, such as strong and variable 
Ca\,{\sc ii} H \& K and 
H$\alpha$ emission \citep{Rodono87,Dempsey96,Robinson96,Garcia-Alvarez02b}. This is due to its tidally 
induced 
rapid rotation, combined with the deepened convection zone of a post-main-sequence envelope. \citet{Ayres01} reported that the K1\,IV component dominates the X-ray emission in the binary system HR 1099.

The G5 dwarf is probably also active, having a sufficiently deep convection zone and 
fast rotation to host an efficient dynamo. In particular, due to the presence of spots, main-sequence G-type 
stars with rotation periods of about 3 days {are expected to} show V-band light curve 
amplitudes up to 0.10 mag, as can 
be inferred from the rotation-activity relations at photospheric levels \citep{Messina01}. However, 
because of the small luminosity ratio {in the V-band} (L$_{\rm G5V}$/L$_{\rm K1IV}\simeq 
0.27$), the contribution 
from the G5 component to the observed magnetic activity manifestations {in the optical band
 is significantly smaller than from the K1\,IV primary.}
 
Extensive broadband photometric monitoring has been a key factor in verifying the presence of evolving spots on the surface of HR 1099 and deducing basic spot parameters (i.e., size, temperature, longitude). HR 1099 is also among the brightest spotted stars and, therefore, a important target for Doppler Imaging. \citet{Vogt83} presented the first Doppler Image of HR 1099. \citet{Gondoin86} used a technique, originally applied to equivalent width (hereafter EW) variations of Ap stars, to map the HR 1099 surface. \citet{Foing94} obtained complete phase coverage for Doppler imaging of spots during the MUSICOS 1989 campaign on HR 1099. \citet{Vogt99} presented a seminal paper including 23 Doppler images obtained from 1981 to 1992. They showed the presence of a long-lived {($>$11 yr)} polar spot, together with transient ($<$1 yr) low-latitude 
spots on the surface of the active K star. \citet{Strassmeier00} reported latitudinal and longitudinal migration of individual starspots and their morphological evolution. They confirmed the poleward migration scenario previously detected by \citet{Vogt96}. We note that all the authors, except \citet{Gondoin86}, have found a polar spot on the surface of HR 1099. Strong magnetic fields
of the order of 1000 G have also been detected 
on HR 1099 using the Zeeman-Doppler Imaging technique 
{\citep{Donati90,Donati92,Donati99}.}

This work follows that done by \citet{Garcia-Alvarez02b} on HR 1099 and based on the MUSICOS 1998 campaign. {Using photometry spot modelling, they obtained maps of the distribution of the spotted regions on the photosphere of the binary components. Those maps were derived using the Maximum Entropy method and Tikhonov regularization criterion on photometric data sets only.} The maps show the K1\,IV primary to be the more active with a large spotted 
region in the northern hemisphere, centered around phase 0.85 (computed according to the ephemeris given in Eq.~(1)). The detection of optical and X-rays flare events around the same phase is in good agreement with a spatial link between flares and 
active regions as reported by \citet{Garcia-Alvarez02b}. In this paper, we present the 
results of the Doppler Imaging technique applied to the spectroscopic data sets obtained at Kitt Peak National Observatory (hereafter KPNO) and Mt. Stromlo Observatory (hereafter MSO), during the MUSICOS 1998 campaign on HR 1099. Observations and data analysis 
are described in Sect.~2; in Sect.~3 we present the image reconstruction process, and in Sect.~4 the results and discussion 
are given. 
\begin{table}[htbp!]
\begin{center}
\caption{\small{The log of observations of HR 1099 and Standards from KPNO and MSO during the MUSICOS 1998 campaign.}}
\scriptsize{
\begin{tabular}{@{}lcrcl@{}}
\hline
\hline\\
 &  & KPNO &  & \\
\hline\\
Object & UT Start & Exp.time & No. of & Comments \\
       & 1998 Nov 24 & (sec.) & frames&\\
\\
HD 218045 & 01:16 & 300 & 2 & B9\,III telluric std.\\
HD 5268   & 01:33 & 300 & 1 & G5\,IV template\\
HR 1099   & 01:53 & 1200 & 4 & Target Star\\
 &  &  &  & \\
Object & UT Start & Exp.time & No. of & Comments \\
       & 1998 Nov 25 & (sec.) & frames&\\
\\
HD 218045 & 01:23 & 300 & 1 & B9\,III telluric std.\\
HD 5268   & 01:32 & 300 & 1 & G5\,IV template\\
HR 1099   & 01:51 & 1200 & 4 & Target Star\\
HD 82074  & 11:44 & 300 & 1 & G6\,IV template\\
HD 92588  & 12:01 & 300 & 1 & K1\,IV template\\
 &  &  &  & \\
Object & UT Start & Exp.time & No. of & Comments \\
       & 1998 Nov 26 & (sec.) & frames&\\
\\
HD 218045 & 05:23 & 300 & 2 & B9\,III telluric std.\\
HR 1099   & 06:03 & 1200 & 5 & Target Star\\
 &  &  &  & \\
Object & UT Start & Exp.time & No. of & Comments \\
       & 1998 Nov 27 & (sec.) & frames&\\
\\
HD 218045 & 02:29 & 300 & 2 & B9\,III telluric std.\\
HR 1099   & 02:56 & 1200 & 7 & Target Star\\\\
\hline
\hline\\
 &  & MSO &  & \\
\hline\\
Object & UT Start & Exp.time & No. of & Comments \\
       & 1998 Nov 29 & (sec.) & frames&\\
\\
HR 1099  & 12:16 & 900 & 3 & Target Star\\
HD 36079 & 16:11 & 300 & 1 & G5\,II template\\
 &  &  &  & \\
Object & UT Start & Exp.time & No. of & Comments \\
       & 1998 Nov 30 & (sec.) & frames&\\
\\
HD 5112  & 09:48 & 300 & 1 & M0\,III template\\
HR 1099  & 10:10 & 900 & 4 & Target Star\\
HD 39523 & 17:51 & 300 & 1 & K1\,III template\\
 &  &  &  & \\
Object & UT Start & Exp.time & No. of & Comments \\
       & 1998 Dec 01 & (sec.) & frames&\\
\\
HD 218045 & 09:43 & 300 & 1 & B9\,III template\\
HD 886    & 09:51 & 300 & 1 & B2\,IV telluric std.\\
HR 1099   & 10:14 & 800 & 4 & Target Star\\
HD 39523  & 17:57 & 300 & 1 & K1\,III template\\
 &  &  &  & \\
Object & UT Start & Exp.time & No. of & Comments \\
       & 1998 Dec 02 & (sec.) & frames&\\
\\
HD 218045 & 09:44 & 300 & 1 & B9\,III template\\
HR 1099   & 10:17 & 900 & 1 & Target Star\\
 &  &  &  & \\
Object & UT Start & Exp.time & No. of & Comments \\
       & 1998 Dec 03 & (sec.) & frames&\\
\\
HD 5112   & 10:55 & 300 & 1 & M0\,III template\\
HR 1099   & 11:16 & 800 & 5 & Target Star\\
HD 36079  & 16:53 & 300 & 1 & G5\,II template\\
\hline \\
\end{tabular}
}
\end{center}
\end{table}
\section{Observations and Datasets}
The analysis of the rotational modulation of the optical bands flux (i.e., U, B, V) allows us to obtain maps of the stellar photosphere or more precisely, temperature (in the case of multiband photometry), total area and longitudinal distribution of the spotted regions. The deeper the minimum in the light curve, the larger the 
spotted region on the stellar disk for a given temperature
difference between the unperturbed photosphere and the spots.
The shape of the light curve is usually not sinusoidal, indicating
that more than one active region is required to interpret the
optical flux modulation \citep[e.g.,][]{Rodono86}. However, apart from eclipsing binaries, no reliable constraints can be derived from photomery for what concerns the latitudinal distribution of spotted regions. Moreover, due to limb-darkening effects only spots located at low latitudes in equator-on stars, or more generally, near the substellar latitudes 
contribute appreciably to the observed light curve modulation.
Thus, while photometry is most sensitive to spots at latitudes much higher/lower  than the substellar ones, the spectroscopy reconstructions have the best latitude distrimination at higher latitudes than substellar ones \citep[e.g.,][]{Unruh97}. Therefore, in order to obtain the 
best possible reconstruction of the distribution of brightness on a 
stellar photosphere it is necessary to combine both spectroscopic and
photometric data sets. 

\subsection{MUSICOS 98 campaign: Optical Spectroscopy}
The data analysed in this work were obtained at two sites, namely: Kitt Peak National Observatory (KPNO), USA and Mt. Stromlo Observatory (MSO), Australia, as part of the MUSICOS 1998 campaign.

The KPNO data set was obtained on 1998 November 24-27 using the coude spectrograph and the 3k$\times$1k \footnotesize{F3KB} \normalsize CCD at the 0.9m coude feed telescope. The wavelength range of the extracted spectra is 5465-7060 \AA, recorded over 23 orders with a resolving power of R=65\,000. A total of 20 spectra of exposure time 1200\,s were obtained, along with spectral standard stars with effective temperatures similar to those of the HR 1099 K1\,IV and G5\,V components.

The MSO data set was obtained on 1998 November 29 - December 3 using the echelle spectrograph on the 74-inch telescope. The 4k$\times$2k \footnotesize{CCD17} \normalsize chip was used. A bluer setting with a wavelength coverage of 4540-6700 \AA, recorded over 43 orders with a resolving power of R=35\,000, was used at this site. A total of 17 spectra of exposure time of 800\,s/900\,s were obtained, together with spectral standard stars with effective temperatures similar to the HR 1099 components.

The details of the observations are recorded in Table~1. A more detailed log of observations of the target HR 1099, during the MUSICOS 1998 campaign, can be found in \citet{Garcia-Alvarez02b}.

The spectra were extracted using the standard reduction procedures in the \footnotesize{IRAF}\normalsize\footnote{\footnotesize{IRAF} \normalsize is distributed by the National Optical 
Astronomy Observatories, which is operated by the Association of Universities 
for Research in Astronomy, Inc., under cooperative agreement with the National
Science Foundation.} packages. Background subtraction and flat-field correction using 
exposures of a tungsten lamp were applied. The wavelength calibration was 
obtained by taking spectra of a Th--Ar lamp. A first-order spline cubic fit to 
approximately 45 lines achieved a nominal wavelength calibration accuracy which ranged 
from 0.06 to 0.11 \AA.  The spectra were normalised by a low-order 
polynomial fit to the observed continuum. Finally, for the spectra affected by water lines, a 
telluric correction was applied. 

To produce a phase-folded light 
curve we adopted 
the ephemeris given by \citet{Garcia-Alvarez02b}:
\begin{equation}
HJD=2451142.943 + 2.83774E 
\label{ephemeris}
\end{equation}

\subsection{Photometry}
During the MUSICOS 98 campaign, contemporary ground-based photoelectric observations were obtained
with two different Automatic Photoelectric
Telescopes (APTs):  a) the 
0.80-m APT-80 of Catania Astrophysical Observatory (CAO)  on Mt.\,Etna, Italy 
\citep{Rodono01}; 
b) the  0.25-m T1 {\it Phoenix} APT of Fairborn Observatory at Washington 
Camp, AZ, USA, that is managed as a multiuser telescope \citep{Boyd84,Seeds95}. Both 
telescopes 
are equipped with standard Johnson UBV filters and uncooled photoelectric 
photometers. Although repeated observations were scheduled on each night during the campaign to obtain complete 
rotation phase coverage, bad weather conditions resulted in light curve gaps around phases 0.30 
and 0.70. Details on the observation and reduction can be found in \citet{Garcia-Alvarez02b}.
\begin{figure*}[htbp!]
   \centering
\begin{turn}{0}
 \includegraphics[width=7cm]{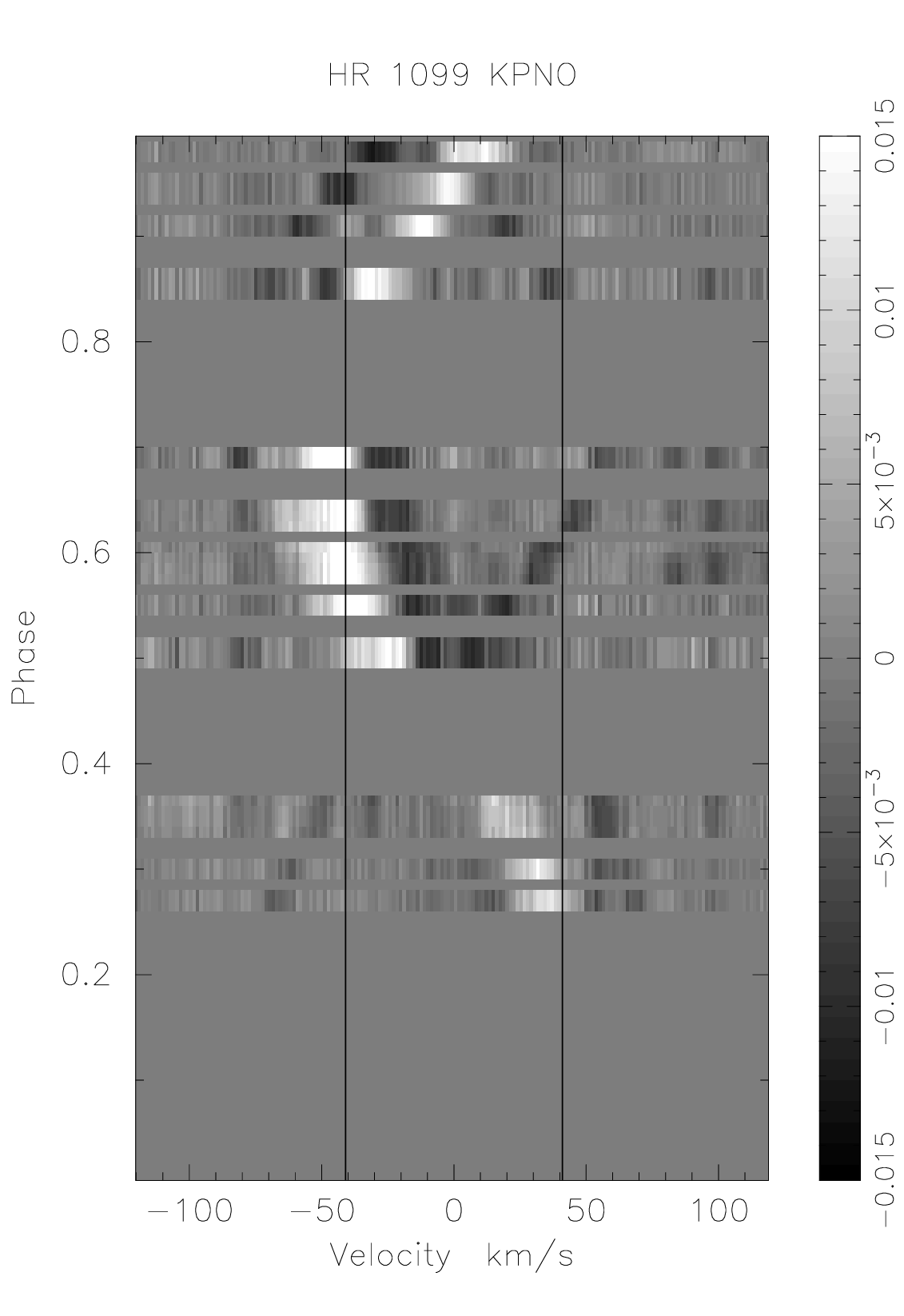}
 \includegraphics[width=7cm]{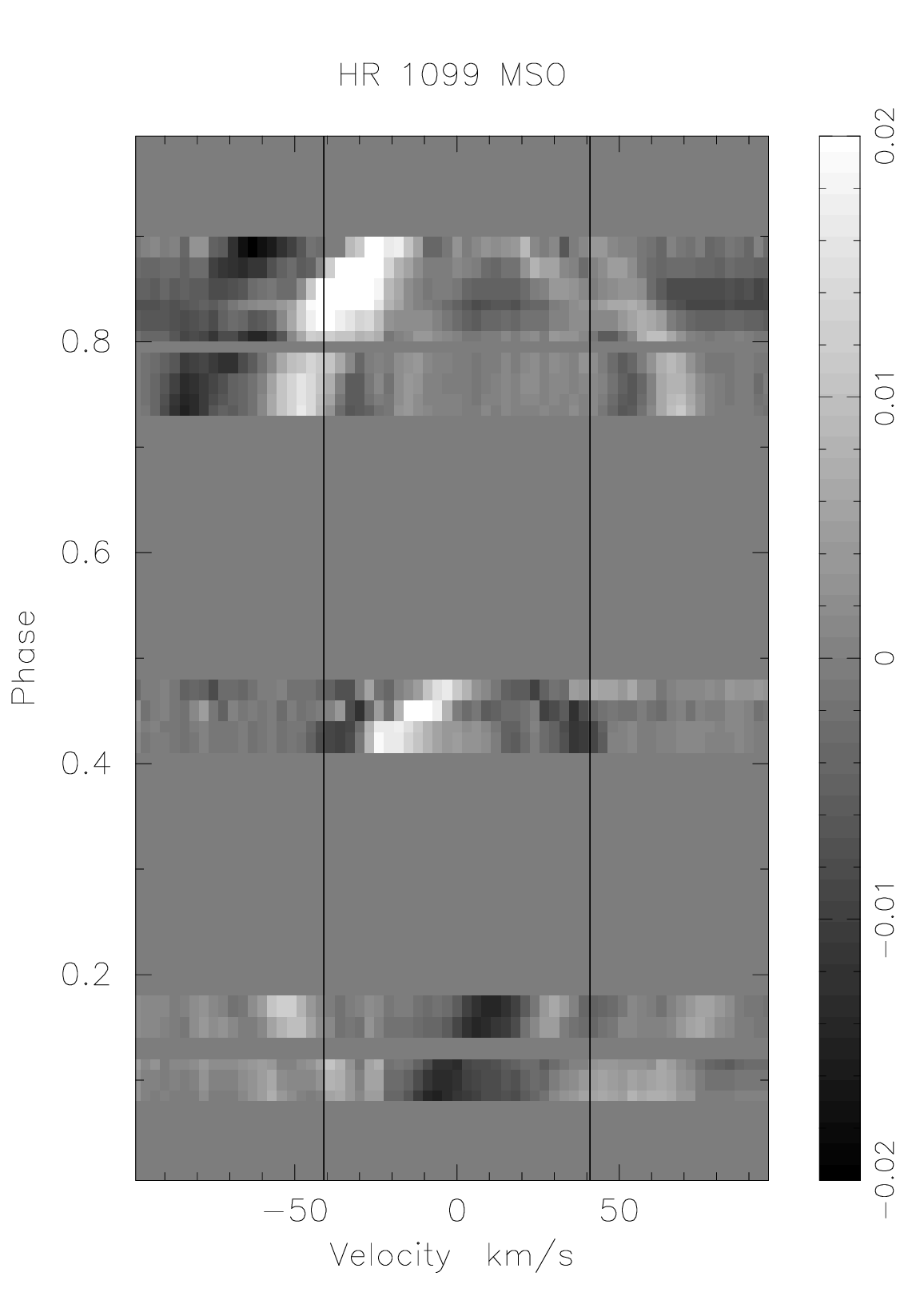}
\end{turn}
   \caption{\small{KPNO (left panel) and MSO (right panel) grey-scale time-series images of the deconvolved spectra of both components of HR 1099, for each site. The plots are residuals after subtraction of the mean profile for each data set (white features are spot transients). The vertical lines are at the $v$\,sin\,$i$ of the K1\,IV component.}}
\label{Time_Series_KNPO_MSO}
\end{figure*}
\begin{table}[tbp!]
\begin{center}
\caption{\small{Least-squares multiplex gain for each data set. The output S/N ratios are calculated from the \footnotesize{LSD}\normalsize  process.}}
\scriptsize{
\begin{tabular}{@{}lcccc@{}}
\hline
\hline\\
Data Set & Input & LSD output & Gain & Effective \\
       & S/N ratio & S/N ratio & S/N ratio& No. of lines\\
\hline\\
KNPO 24th  & 65 & 1790 & 27.7 & 1711\\
KNPO 25th  & 73 & 2038 & 27.9 & 1711\\
KNPO 26th  & 74 & 2052 & 27.8 & 1711\\
KNPO 27th  & 78 & 2163 & 27.9 & 1711\\
MSO 29th  &  86& 3656 & 42.3 & 2249\\
MSO 30th  &  92& 3861 & 42.2 & 2249\\
MSO 01th  &  83& 3539 & 42.4 & 2249\\
MSO 02th  &  59& 2489 & 41.9 & 2249\\
MSO 03th  &  87& 3634 & 41.6 & 2249\\
\hline \\
\end{tabular}
}
\end{center}
\end{table}

\section{Image Reconstruction}

In order to obtain the best possible results, the Doppler imaging technique requires careful preparation of the observations. This section describes the steps that have been followed to reach this aim.

\subsection{Continuum fitting}
The best possible fit to the observed continuum is required, in order to obtain the best results from the least-squares deconvolution. The continuum of an echelle spectrum can be estimated by using several methods. The blaze function dominates the shape of the continuum in the echelle orders.
For CCDs of sizes of 2k$\times$2k pixels, or even larger, the continuum is generally well approximated by a carefully chosen spline fit. The number of knot points required in the spline fit varies from one data set to the next. In this case, eight knot points were used for standard stars on both KPNO and MSO data sets, while eleven knot points were used for the target star, HR 1099. 
\begin{figure*}[htbp!]
   \centering
\begin{turn}{0}
\hspace{2.cm}
 \includegraphics[width=12.5cm]{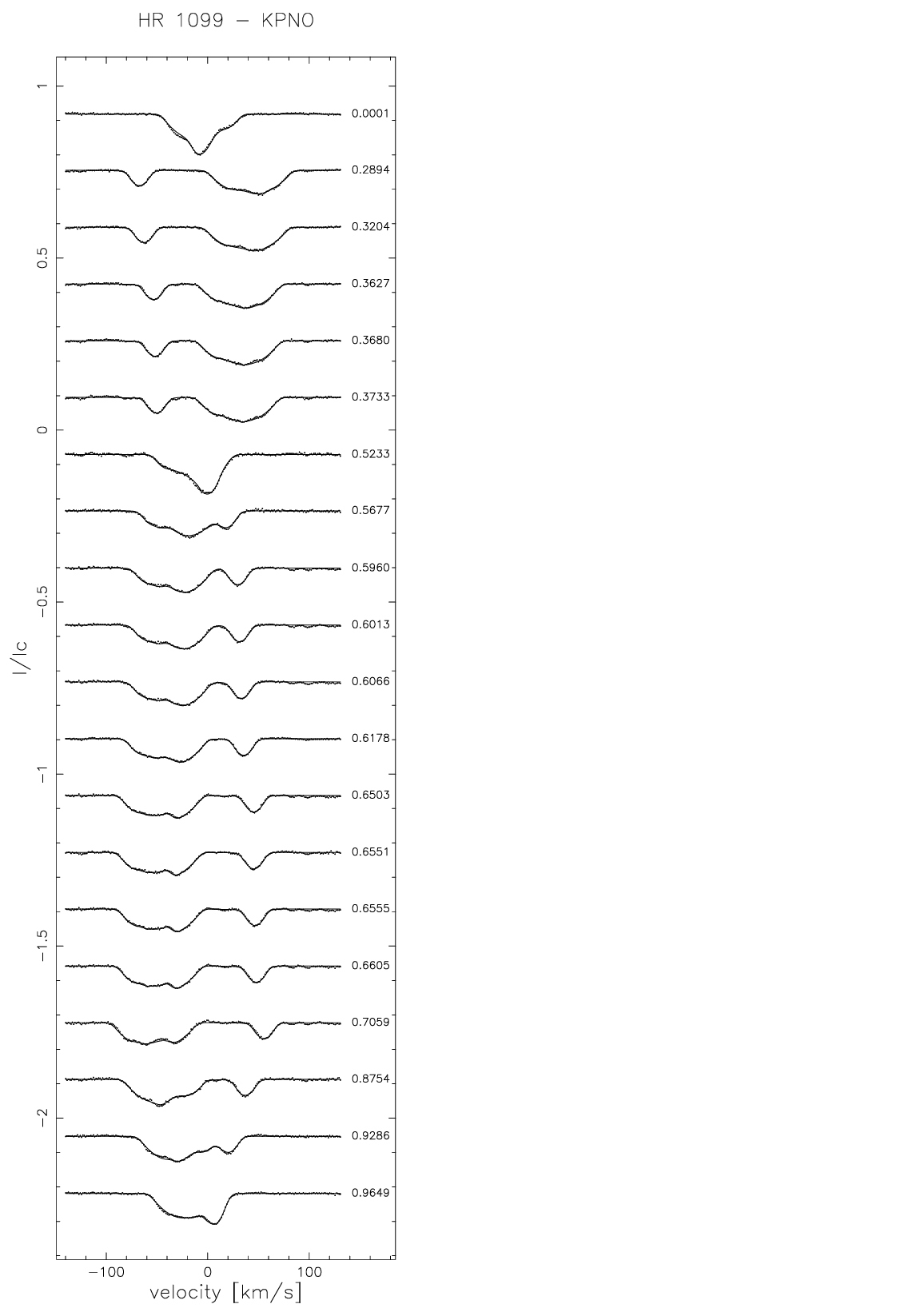}
\hspace{-6cm}
 \includegraphics[width=12.5cm]{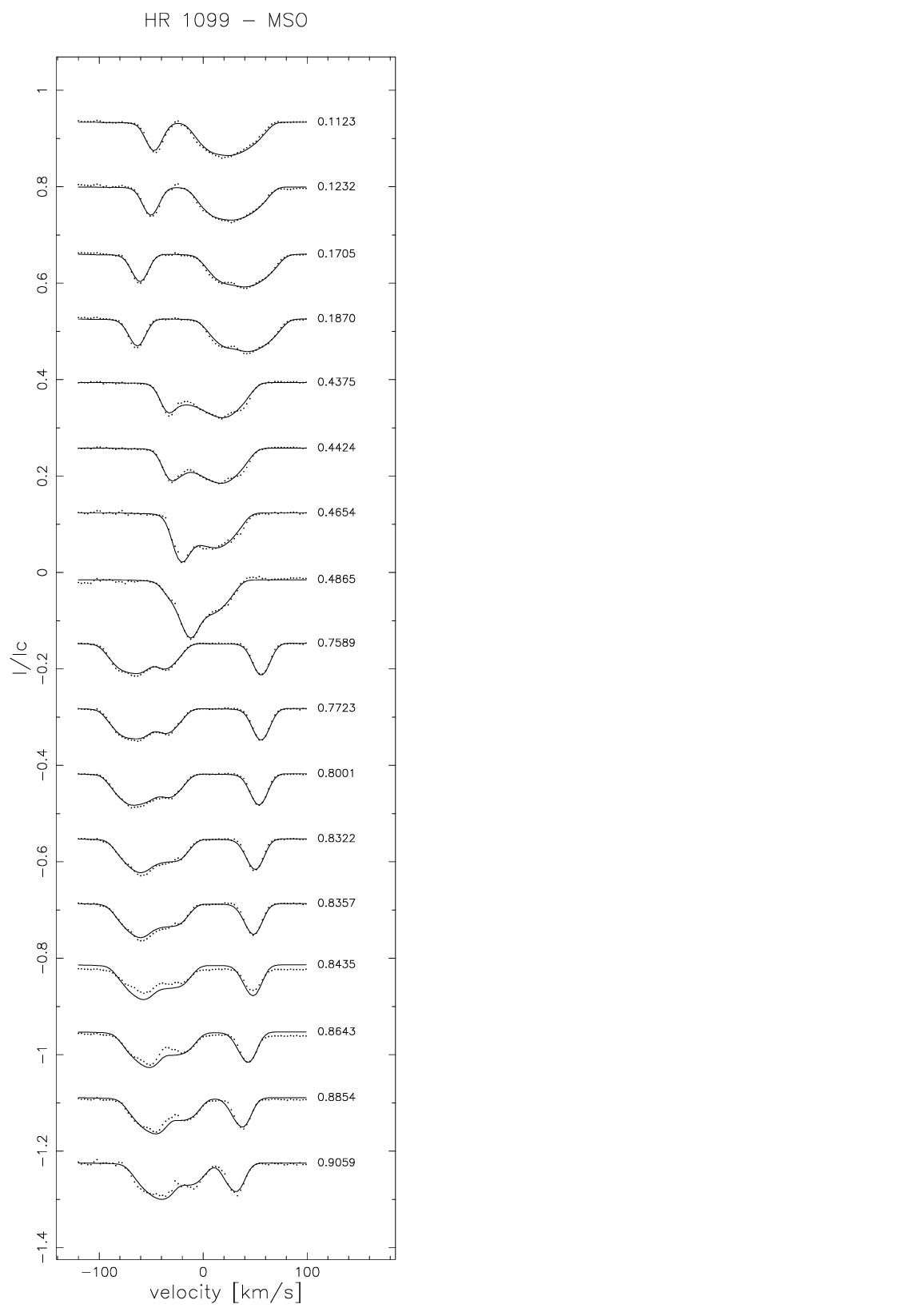}
\end{turn}
   \caption{\small{Time-series profiles and maximum-entropy-regularized fits to the HR 1099 binary system data observed at KPNO (left panel) and MSO (right panel) during the MUSICOS 1998 campaign. The number to the right of each profile is the phase of the observation. We have plotted the spectra
   ordered in phase for easier comparison between them. The level of the continuum has been systematically shifted to simplify the comparison of all the profiles.}}
   \label{FigMEMSpectro_KNPO_MSO}
\end{figure*}

We have taken as a master frame a spectrum in which one component of the system is in front of the other. This allows us to observe the largest amount of continuum possible. This master continuum was then divided by the fit to the continuum of the two slowly rotating standards that match the spectral types of the components of the HR 1099 system. Note that different standards with different spectral types were used for the KPNO and MSO data sets, respectively. This continuum fitting process is similar to that first described by \citet{Collier-Cameron94}. This approach has the advantage that the lines do not suffer from rotational broadening, giving a greater number of continuum windows. Since the continuum level and shape of HR 1099 differ slightly from those of the photospheric standards, a low order (quartic) polynomial was fitted to this divided spectrum to remove residual misfits, and to normalise the spectrum, resulting in a standard or master continuum fit. 

Normalisation of the HR 1099 spectra were carried out by dividing each target spectrum by the master continuum frame, in order to obtain an essentially flat residual. This residual was then fitted with a quartic polynomial to remove residual misfits and any remaining tilt, mainly due to extinction effects as the star rises and sets. Slight changes in continuum tilt due to the starspot contribution are also removed in the process. This fit, when multiplied by the master continuum fit, results in the correct continuum for the target spectrum.

\subsection{The local specific intensity profiles.}
A spotted star is considered to have two temperature components (i.e. photospheric temperature and spot temperature). In the case of a binary system, two photospheric temperatures and two spot temperatures, for the primary and secondary component respectively, are needed. Two template stellar spectra to mimic the photosphere temperature were used, namely: a K1\,IV (HD\,92588 and HD\,39523 for the KPNO and MSO data set, respectively) with a photospheric temperature $T$=4\,800\,K for the primary component, and a G5\,V (HD\,5268 and HD\,36079 for the KPNO and MSO data set, respectively) with temperature  $T$=5\,400\,K for the secondary component. The template stellar spectra M0\,III (HD 5112), with $T$=3\,800\,K, was used for spot temperature in both components, in the MSO data set. The KPNO data set does not contain M type standard, so to mimic the spot temperature we used an artificial echelle spectrum generated by using a $T$=3\,800\,K line list \citep[Vienna Atomic Line Database;][]{Kupka99}. {The choice of starspots
with temperature $T$=3\,800\,K is based upon similar single stars or RS CVn subgiants for which the
spot umbrae is ~500-1\,500\,K cooler than the photosphere. These values were obtained either by using multicolour photometry \citep[i.e.,][]{Olah02,Olah97} or by using TiO absorption bands \citep[i.e.,][]{Saar01,Neff95}. We used the same spot temperature in both binary components in order to simplify the analysis. However, if the spot temperature is wrong, incorrect limb-darkening coefficients
will be used but, because of the small effective intensity of the
spot (typically less than 1/10th the photospheric contribution), this will
only affect the spot filling factor obtained in the images rather than the shape of any spot features.} There are indications that the spot properties are sensitive to surface gravity, therefore, when the M0\,III spectra is used as template, for spots on the components of HR 1099, we should take into account that we are working with a certain level of approximation. 

M dwarf spectra contain molecular absorption bands, whose opacities are poorly represented in most available model atmospheres. However, \citet{Barnes01} concluded that the use of an incorrect line list does not severely affect the shape of the deconvolved profile. All the template stellar spectra were deconvolved, using the method of least-squares deconvolution, in the same manner and using the same scaled continuum frame, ensuring correct equivalent width.

\subsection{Limb-darkening coefficient}
The centroidal wavelength was 6043 \AA\ and 5423 \AA\ for KNPO and MSO data sets, respectively. It was calculated from lines in the line list, taking into account both the strength and the variance from the standard fit used in the spectral deconvolution procedure (hereafter \small{SPDECON}\normalsize). Due to the fact that the limb-darkening coefficient is a fairly linear function of wavelength, we used linear interpolation to obtain a value from local thermodynamic equilibrium (LTE) models \citep{Diaz-Cordoves95,Claret95}. The limb-darkening coefficients for the primary component K1\,IV are 0.761 and 0.729 for the 3\,500\,K spot temperature and for the 4\,750\,K photospheric temperature respectively. For the secondary component G5\,V the limb-darkening coefficients are 0.710 and 0.646 for the 3\,600\,K spot temperature and for the 5\,500\,K photospheric temperature respectively. In this analysis 10 different limb angles (i.e., $\mu$=0.1,0.2,...,1.0) for each temperature were used.  
\begin{figure*}[htbp!]
   \centering
\hspace*{-2.cm}
\begin{turn}{270}
\vspace{-3.5cm}
\includegraphics[width=12.5cm]{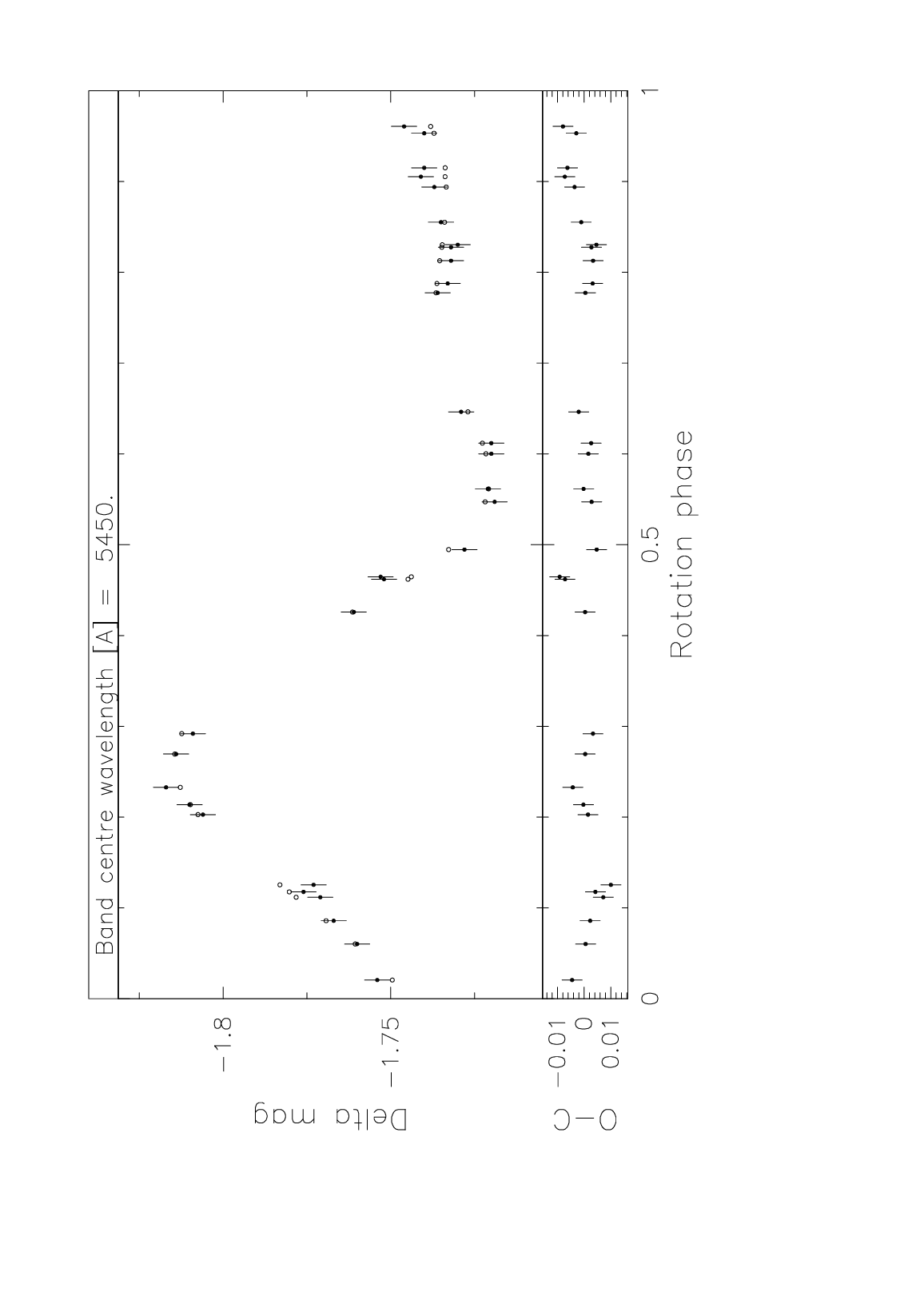}
\end{turn}
\vspace*{-2.5cm}
   \caption{\small{Photometric light curve of HR 1099 obtained contemporary to the MUSICOS 98 campaign. Top panel: Filled dots represent differential V-band  magnitudes and open dots the maximum entropy regularized fit. Bottom panel: Residuals.}}
   \label{FigMEMPhotom}
\end{figure*}
\begin{table*}[htbp!]
\begin{center}
\caption{\small{Estimated system parameters from maximum-entropy minimized $\chi^{2}$ fits to data from each site. $\gamma$ denote the systemic velocity, $K_{\rm 1}$ is the velocity amplitude for the primary component, EW$_{\rm pri}$ and EW$_{\rm sec}$ are the equivalent width of the primary and secondary components respectively, $M_2/M_1$ is the mass ratio, i is the inclination angle and e is the eccentricity.}}
\scriptsize{
\begin{tabular}{lcccccccccc}
\hline
\hline\\
Site & $\gamma$ &$K_{\rm 1}$ &EW$_{\rm pri}$ & EW$_{\rm sec}$ & Period & R$_{\rm pri}$& R$_{\rm sec}$&  $M_2/M_1$     & i &e\\
     & (km\,s$^{-1}$)&(km\,s$^{-1}$) & (\AA) & (\AA) & (d) & (R$_{\odot}$)& (R$_{\odot}$)                   &  &(deg)&\\
\hline\\
KPNO& -12.74& 49.66&117& 139&2.8377&3.3&1.1 & 0.789 &40&0\\
MSO & -15.69& 49.66&82 &210 &2.8377&3.3&1.1&  0.789 &40&0\\
\hline \\
\end{tabular}
}
\end{center}
\end{table*}

\subsection{Least-squares deconvolution}
We use the technique of least-squares deconvolution (\footnotesize{LSD} \normalsize hereafter), first implemented by \citet{Donati97}, to derive an average line profile. This technique can be used to combine the information content of many weak and intermediate-strength photospheric absorption lines, by computing average profiles from thousands of spectral echellogram lines simultaneously. We assume that an echelle spectrum contains the same starspot signatures in all the photospheric Doppler-broadened profiles. Although the amplitude of the signature may change for different lines, the morphology should remain the same. A detailed mathematical explanation of the \footnotesize{LSD} \normalsize technique can be found in \citet{Donati97} and \citet{Collier-Cameron01}.

The maximum achievable gain using \footnotesize{LSD} \normalsize is approximately (number of lines)$^{1/2}$, however this is an upper limit not reachable in most of the cases. In our KPNO and MSO data sets $\sim$1711 and $\sim$2249 photospheric absorption lines are available, respectively. Table~2 shows the S/N obtained by our deconvolution code \footnotesize{SPDECON} \normalsize along with other statistics. The velocity bin width in the deconvolved profiles is generally set to match the mean velocity bin size of each pixel on the CCD. In the case of the KPNO data set the deconvolved profiles for all the HR 1099 observations were set to a bin width of 1.3\,km\,s$^{-1}$, while for MSO the value was set to 3.0\,km\,s$^{-1}$.

Lines in the regions of known strong TiO bands (such as at 6782, 7055 and 7126 \AA) were removed from the line list. Lines with wavelength $>$7200 \AA\ are not included in the deconvolution due to the presence of numerous telluric lines. All the regions around the strongest atomic lines such as H$\alpha$ (6555-6570 \AA), H$\beta$ (4840-4880 \AA), Na Doublet (5880-5910 \AA) and other lines that can be affected during flares such as He~{\sc i} D$_{3}$ (5870-5880 \AA) and Mg~{\sc i} (5160-5190 \AA), were also removed from the line list. 

The time-series spectra obtained after applying least-squares deconvolution are shown in Fig.~\ref{Time_Series_KNPO_MSO}. In the time-series spectra it is possible to identify a very clear starspot signature, visible in both data sets. This feature seems to reaches its maximum around phases 0.6 and 0.7, which is close to the phase of flux minimum in the photometric data set. We also note that, between phase 0.1 and 0.2, there is an apparent lack of starspots, which is in agreement with the maximum in the photometric light curve around those phases. 
\begin{figure*}[htbp!]
\hspace*{-1.cm}
   \centering
\begin{turn}{270}
 \includegraphics[width=11cm]{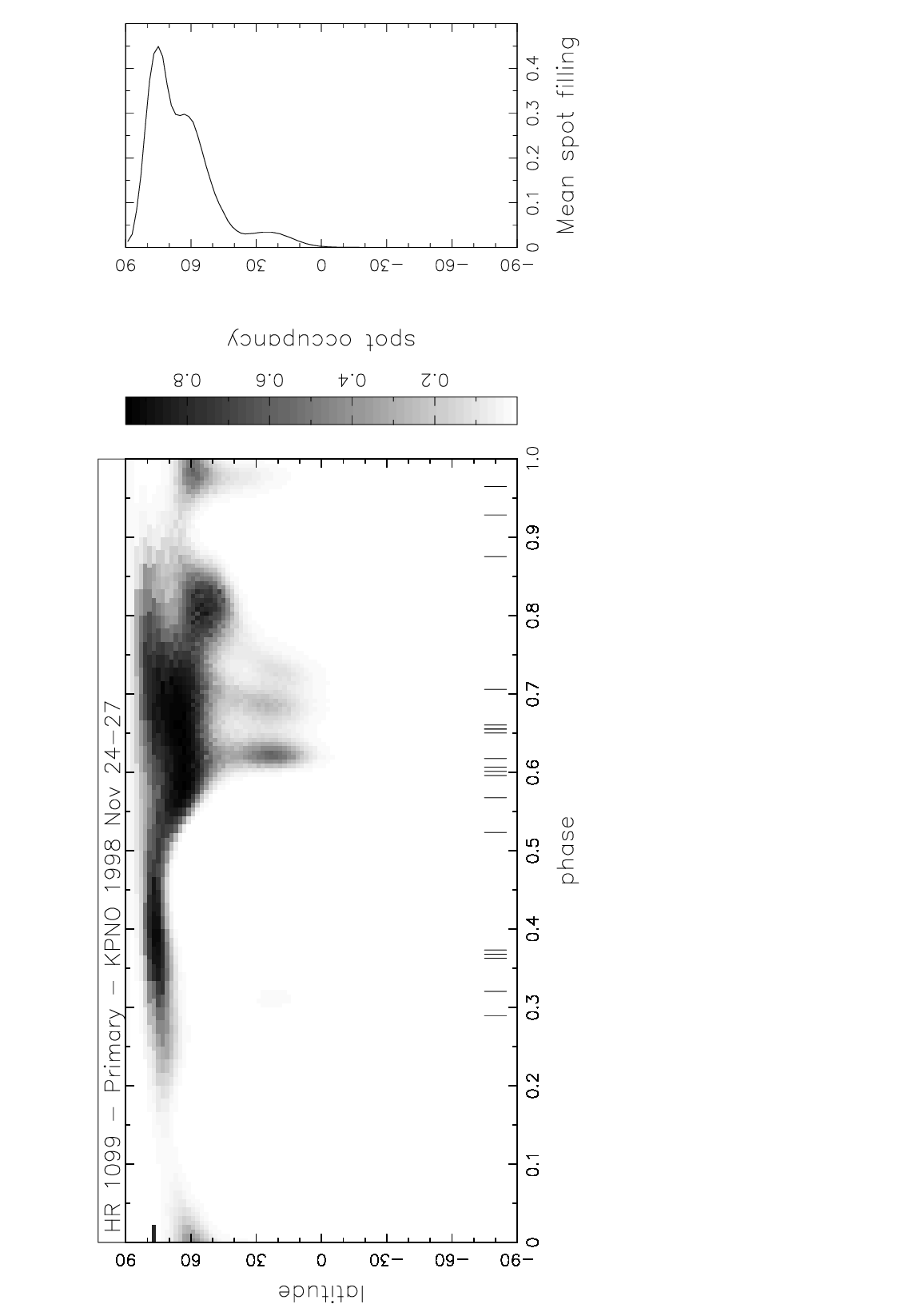}
\end{turn}
\vspace*{-3.5cm}
   \caption{\small{Maximum-entropy-regularized image reconstructions, for the primary component (K1\,IV) of the HR 1099 system, from KPNO data. The tick marks indicate the phases of observation. Note that the longitude runs in the opposite sense to phase, with longitude 0$^{\circ}$ at phase 1.0 and longitude 360$^{\circ}$ at phase 0.0. Right-hand panel: The mean fractional spot occupancy of the reconstructed image as a function of latitude.}}
   \label{FigDIPrimaryKPNO}
\end{figure*}
\begin{figure*}[htbp!]
\hspace*{-1.cm}
   \centering
\begin{turn}{270}
 \includegraphics[width=11cm]{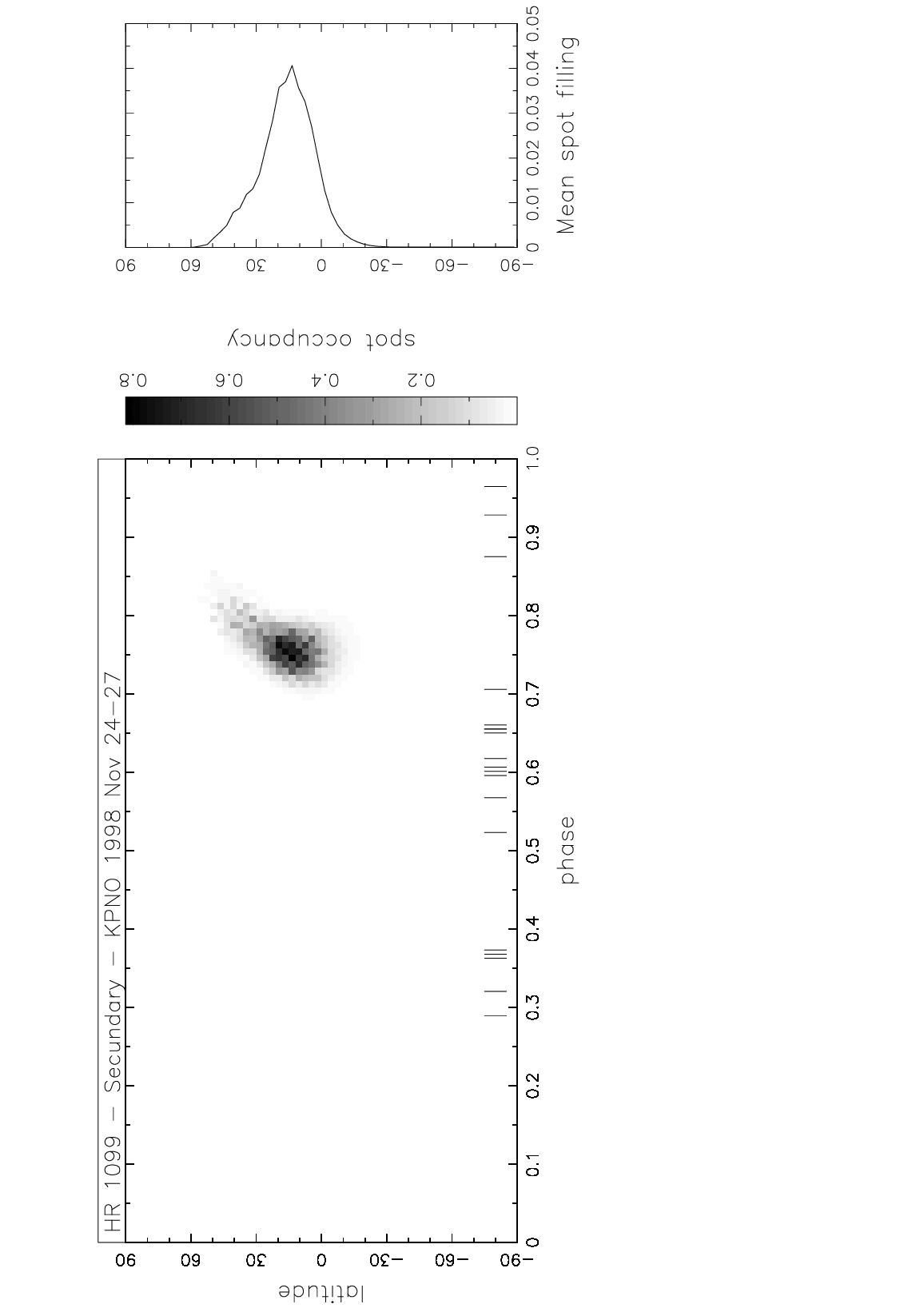}
\end{turn}
\vspace*{-3.5cm}
   \caption{\small{Same as Fig.~\ref{FigDIPrimaryKPNO}, but for the secondary component (G5\,V) of the HR 1099 system.}}
   \label{FigDISecondaryKPNO}
\end{figure*}
\begin{figure*}[htbp!]
\hspace*{-1.cm}
   \centering
\begin{turn}{270}
 \includegraphics[width=11cm]{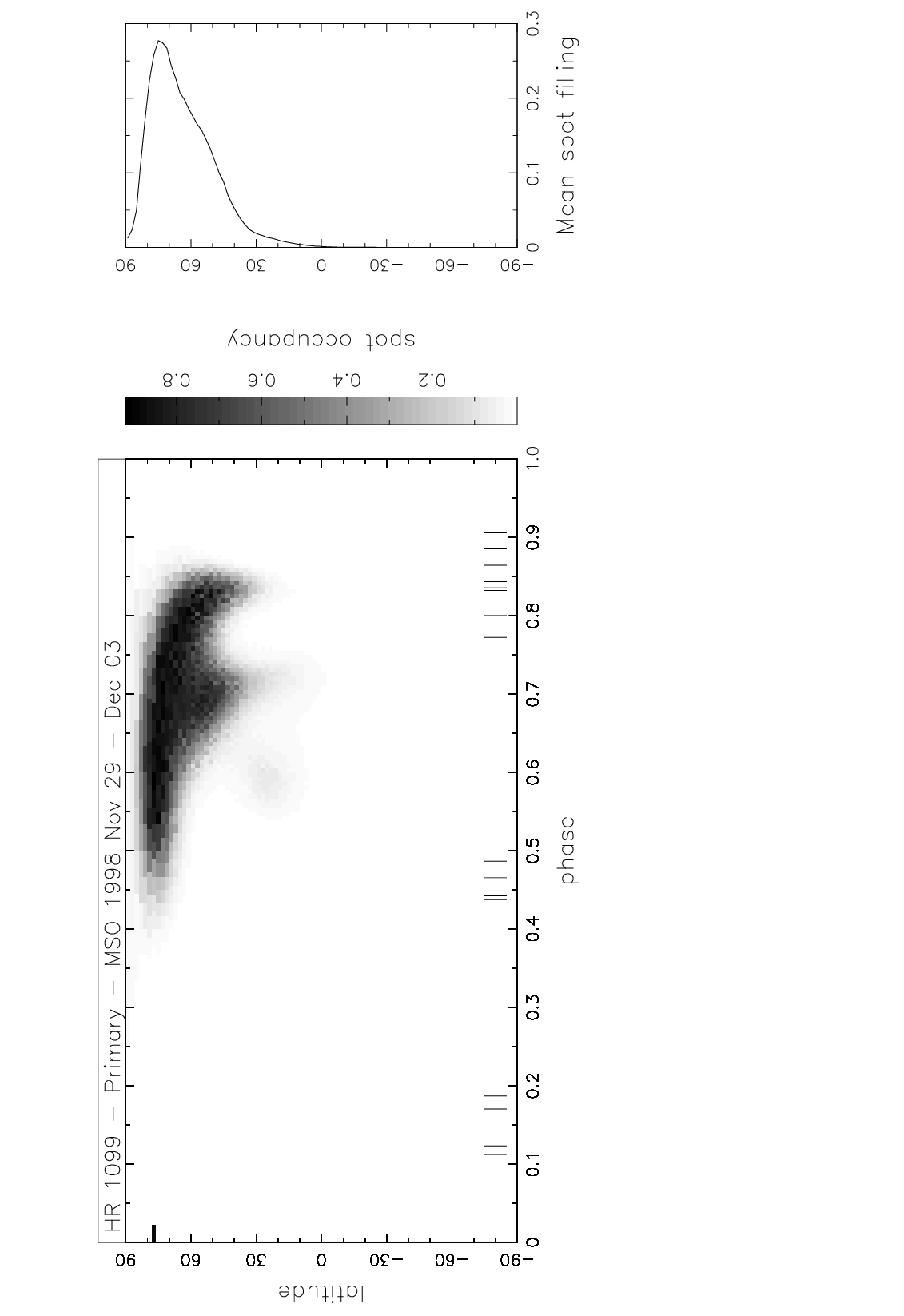}
\end{turn}
\vspace*{-3.5cm}
   \caption{\small{Maximum-entropy-regularized image reconstructions, for the primary component (K1\,IV) of the HR 1099 system, from MSO data. The tick marks indicate the phases of observation. Note that the longitude runs in the opposite sense to phase, with longitude 0$^{\circ}$ at phase 1.0 and longitude 360$^{\circ}$ at phase 0.0. Right-hand panel: The mean fractional spot occupancy of the reconstructed image as a function of latitude.}}
   \label{FigDIPrimaryMSO}
\end{figure*}
\begin{figure*}[htbp!]
\hspace*{-1.cm}
   \centering
\begin{turn}{270}
 \includegraphics[width=11cm]{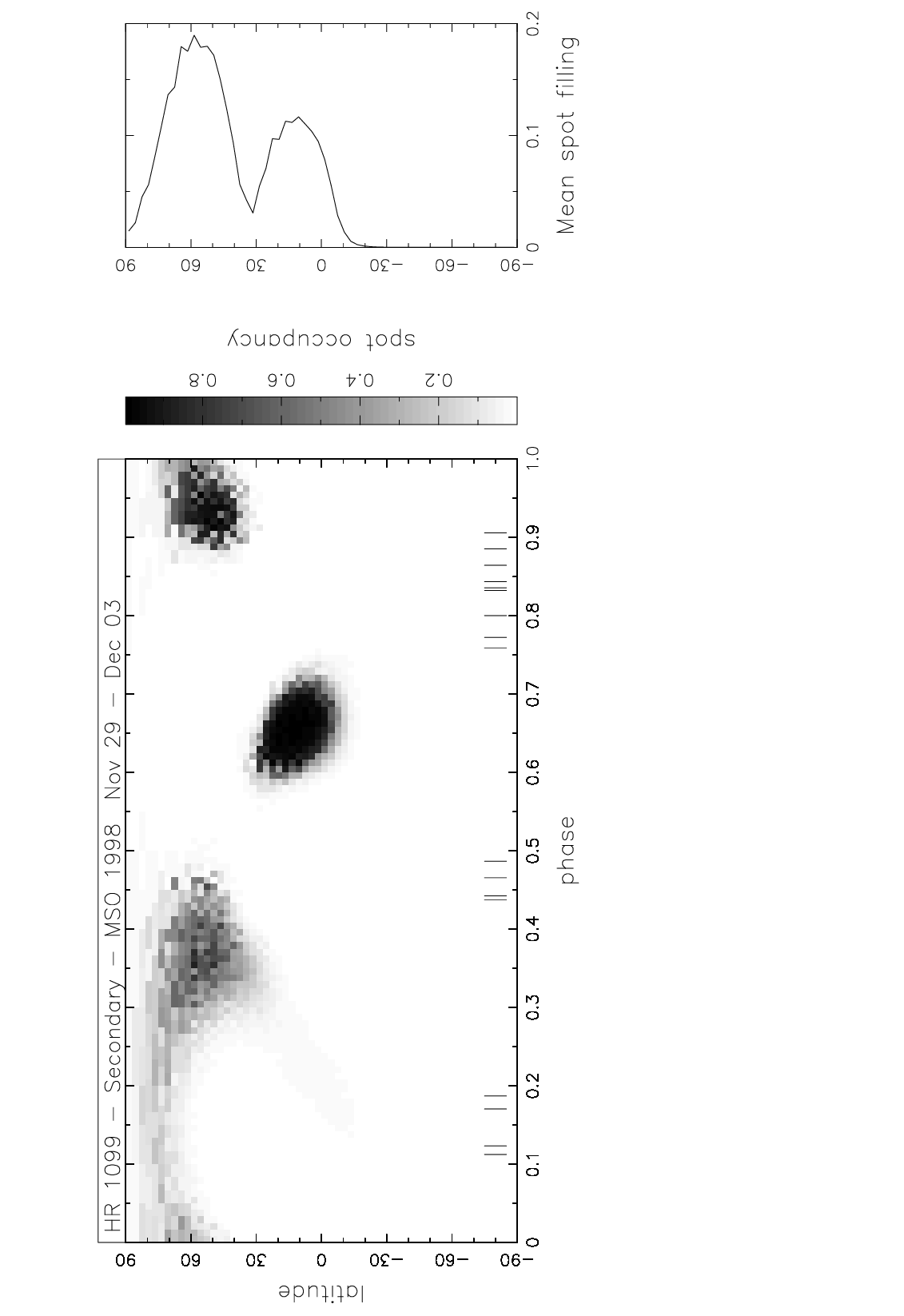}
\end{turn}
\vspace*{-3.5cm}
   \caption{\small{Same as Fig.~\ref{FigDIPrimaryMSO}, but for the secondary component (G5\,V) of the HR 1099 system.}}
   \label{FigDISecondaryMSO}
\end{figure*}

\subsection{Determination of system parameters}
Accurate system parameters, such as the systemic velocity, equivalent width, inclination angle, etc, are needed to obtain the best results from Doppler Imaging. In fact, spurious features are produced as a result of incorrect input parameters values \citep{Collier-Cameron94}. System parameters can be determined empirically by minimization of either spot area (i.e., in the reconstructed images) or minimization of the $\chi^{2}$ fit to the data. We have chosen the $\chi^{2}$ minimization method, since this is more satisfactory for spectra of high S/N ratios and does not introduce problems as the spot area method does \citep{Barnes00}. The $\chi^{2}$ method works as follow: After obtaining initial estimates for the input parameters, we keep all them fixed, except one. A large number of maximum entropy regularised iterations are then performed by our Doppler-Imaging code \footnotesize{DoTS} \normalsize, in order to converge to a $\chi^{2}$ minimum. The initial parameters were taken from \citet{Garcia-Alvarez02b}. Table~3 shows the derived system parameters. There is good agreement between the values derived from the KPNO data set, and those independently derived from the MSO data set. Only the value of the axial inclination is poorly known. There have been several attempts to measure it, yielding estimates ranging from $\sim$33$^{\circ}$ \citep{Fekel83} to $\sim$45$^{\circ}$ \citep{Donati99}. We choose $i$=40$^{\circ}$ since it gives the minimum chi-squares. Errors of $\pm$10$^{\circ}$ generally have an effect only on the spot filling factor as a function of latitude \citep{Barnes99}. However, very strong polar features could appear by overestimating the axial inclination, under-representing low-latitude features as well. The opposite is true for underestimation of the axial inclination.

Note that the internal shift of the instruments, which could affect the radial velocity measurement, are corrected by using telluric features \citep[see][]{Collier-Cameron99}. This ensures reasonably accurate radial velocity determinations. Taking into account the different phase ranges and different instruments used in both data sets, one would expect some spurious scatter in the radial velocity.

\subsection{Final Image reconstruction}
Surface images of HR 1099 were obtained from the spectroscopic deconvolved line profiles and photometric light curves using the \footnotesize{DoTS} \normalsize surface imaging code \citep{Collier-Cameron97}. Based on the \footnotesize{MEMSYS} \normalsize algorithm developed by \citet{Skilling84}, \footnotesize{DoTS} \normalsize is a maximum entropy code which is used for the Doppler tomography of stellar surfaces. \citet{Collier-Cameron97} describes the stellar surface model and the integration scheme.

Although some authors \citep[e.g.][]{Donati99} have neglected the contribution of the G5\,V secondary component to the global activity of the system, we do try to obtain an image from both components of the binary system. The maximum entropy fits to the KPNO and MSO spectroscopic data sets are shown in Fig.~\ref{FigMEMSpectro_KNPO_MSO}. We observe that the deconvolved profile of the system HR 1099, is well fitted at almost all the orbital phases. However, for phases near 0.0 and 0.5 the fit is not as good due to the degeneracy introduced by the overlapping profiles. Fig.~\ref{FigMEMPhotom} shows the maximum entropy fit to the photometric light curve. Figs.~\ref{FigDIPrimaryKPNO}-\ref{FigDIPrimaryMSO} shows a clear high-latitude spot between phases 0.55 and 0.85.  

The maximum entropy reconstructed images for both KPNO and MSO data sets, Figs.~\ref{FigDIPrimaryKPNO}-\ref{FigDISecondaryMSO}, are a combination of the spectroscopic and photometric data sets. Both KPNO and MSO data sets have been treated independently. A stellar surface grid with 90 latitude bands was used for the image reconstructions.

As we have already said in $\S$3, the spectroscopic reconstructions have poor latitude discrimination at latitude $<$30$^{0}$. On the other hand, the photometric data is sensitive to structures situated at low latitudes, although at much lower resolution. All this implies that we have to weight the spectroscopic and photometric data sets in order to combine them properly. The \footnotesize{DoTS} \normalsize surface imaging code includes a weighting factor, $\beta$, that determines which data set, spectroscopic or photometric, will contribute most to the image reconstruction. The weighing factor can be set from 0 (100\% photometric data) to 1 (100\% spectroscopic data). However, $\beta$ is larger than 0.9 in most of the cases, due to the higher amount of information contained in the spectroscopic data. In our case, the value for $\beta$ was set to 0.90 and 0.92 for the KPNO and MSO data sets, respectively.

\section{Results and Discussion}
A maximum entropy reconstruction of the spot distribution on both components of the binary system HR 1099 is shown in Figs.~\ref{FigDIPrimaryKPNO}-\ref{FigDISecondaryMSO}. The mean fractional spot area as a function of latitude is also plotted for each map. It shows that medium-high latitudes are the main regions of spot coverage for the primary component. The secondary component shows a medium-low latitude spot around orbital phase 0.7. However, the are also other features on the secondary component which are not reliable as we will explain below. 

There seems to be a global similarity between the maximum entropy reconstructed images for the primary, (Figs.~\ref{FigDIPrimaryKPNO} and \ref{FigDIPrimaryMSO}),  where starspots are mainly observed at latitudes higher than 30$^{\circ}$. In both maps the surface appears to be heavily spotted at the phases 0.6 to 0.9, while spotted at the phases 0.2 to 0.6. {All these spots, detected from KPNO and MSO spectroscopic data sets, are in agreement with those detected from photometric spot modelling, using contemporaneous photometry, by \citet{Garcia-Alvarez02b}. These authors, based on spectroscopic data sets obtained from several sites during the MUSICOS 1998 campaign, also observed rotational modulation and chromospheric flares occurring at the same phases.} 

Although our sampling does not cover all the orbital phases, the low inclination of the stellar rotation axis (40$^{\circ}$), indicates that most spot features that appear at high latitudes are in view all the time, contributing to every line profile, independently from the phase. Both Figs.~\ref{FigDIPrimaryKPNO} and \ref{FigDIPrimaryMSO} recover a structured polar spot which is off-centered, with respect to the stellar rotation pole. \citet{Donati03} reported very similar spot distribution on the primary component, from observations around the same epoch, reconstructed from Zeeman-Doppler imaging spectropolarimetry. This near-polar spot was also found by other authors \citep{Vogt99,Donati99,Strassmeier00} by using independent codes. \citet{Hatzes96} presented a detailed study of the reality of polar spots. Their modelling simulation argues strongly against gravity darkening, differential rotation, limb brightening, equatorial bright bands, chromospheric line filling, and unknown effects in line radiative transfer physics as possible causes of the spurious polar spots. They suggested an inclination dependence of the flattening of the core of the line profiles in RS CVn's as an evidence for the reality of the large cool polar spots. The long vertical feature observed in Fig.~\ref{FigDIPrimaryKPNO} at phase 0.6 may be real, because the sampling around this phase is large enough. Although the entropy criterion expresses the uncertainty in latitude by smearing the image north-south at low latitudes, one has to bear in mind that the  combination of the spectroscopic and photometric data sets obtaining the maps should have partly removed this problem. No low-latitude features were obtained in the reconstructed images.  

There are, however differences between the maps, which may be partly due to the different spectral resolution of the KPNO and MSO spectrographs. The higher resolution of the KPNO data (R=65\,000) allow us to detect isolated features and groups of features, while in the lower resolution MSO data set (R=35\,000) those features will appear as a single big spot (e.g., features at latitude 60$^{\circ}$ and phases 0.7-0.9). The resolution problem should not be affected by the input S/N, because it is similar for both KPNO and MSO data sets. Note that we are confident that the data has not been over-fitted. This additional fact could add spurious surface features. We have to take into account that the reduction process is never perfect, combined with incomplete phase coverage. However, by stopping the iterations before the noise is being fitted we ensure a reliable result.

The images for the secondary components obtained from both KPNO and MSO data sets show a few medium-low latitudes features. We observe a low-latitude spot around orbital phase 0.7 in the images from both data sets. This spot seems to mirror the structure seen on the primary component. {However, one has to bear in mind that most of the evidence for the structure on the G5\,V component comes from changes in the EW of the star's contribution around the orbit, since its profile is so narrow. We performed a series of numerical experiments on our HR 1099 data sets, altering the effective temperature of the spots, in order to see if the spot distribution on the secondary is sensitive to the spot temperature assumed. We conclude that the spot observed on the secondary component is not affected by changes in the surface brightness of the spots on the primary. This fact confirm the reliability of our results, insofar as sensitivity of the image to the spot temperature is concerned. However, this low latitude feature observed on the secondary component, is in disagreement with the convective overshoot dynamo models reported by \citet{Granzer00}, which predict the emergence of magnetic field at higher latitudes. We also observe two features which only appear in the MSO data set. They occur at phases where no observations were made and may be ghost features due to a poorly sampled data set \citep{Stout-Batalha99}.}  

{The mirroring effect, observed on the HR 1099 binary system, might suggest that tidal forces may influence the spot distribution. However, our observational data set only covers a few orbital rotations and, in order to claim this mirroring as a real physical effect, it would be better to confirm it by observing the HR 1099 system during consecutive years or even try to find similar behaviours in other binary systems. \citet{Hatzes99} reported the surface images on both components of the active pre-main-sequence binary V824 Arae (HD 155555) obtaining mirror images, as it happens in our observations. Their images look nearly identical if one reverses or flips one with respect to the other. \citet{Piskunov96} and \citet{Piskunov01} suggested that the presence of spots at sub-stellar points, obtained through Doppler imaging, facing the other component on the binary system ER Vul, could be attribute to the reflection effect. This is unlikely to occur in our case considering the separation of the two components of HR 1099, $\sim$11\,R$_{\sun}$. Even more, the maps obtained for the secondary component only show a small spot (consistent in both images) compared with the much larger structure obtained for the primary component. \citet{Gunn97b} presented extreme ultraviolet and radio observations of the RS CVn system CF Tucanae. They detected a modulation in the EUV and radio flux levels which led them to suggest the existence of an intra-binary region of activity. \citet{Gunn99} found a similar region on the Algol system V505 Sgr. These intra-binary regions would imply a joint magnetosphere and possible interaction between magnetic loops between the two components of the binary system. \citet{Gunn99} suggested that their observations could alternatively be interpreted as a surface feature on the later type star facing the companion.}

\begin{acknowledgements}
We wish to thank all those who have contributed to the MUSICOS 98 campaign. 
Research at Armagh Observatory is grant-aided by the Department of Culture, 
Arts and Leisure for Northern Ireland. DGA wishes to thank School of Physics and Astronomy for financial support while visiting St Andrews University. We thank the referee, Rachel Osten, for helpful comments on the manuscript.
\end{acknowledgements}

\bibliographystyle{apj} 
\bibliography{references}

\end{document}